# Non Linear System for a Veritable pid Substitute

**Prof. PhD Petre Bucur,**
**Assoc. Prof. PhD Lucian Luca,**
**Faculty of Computers and Applied Computer Science**
**„Tibiscus" University of Timişoara, România**

**REZUMAT**. Lucrarea se referă la un sistem neliniar folosit foarte mult în biologie, dar care, în anumite condiţii, cu anumite valori ale coeficienţilor devine liniar şi cu o diagramă absolut liniară pe o durată mare de timp. El poate fi folosit ca un regulator veritabil în controlul sistemelor.

## 1 Nonlinear system Lotka-Voltera

Non linear system Lotka Voltera described process of prey-predator evolution in nature. One of them is formulated in normal Cauchy form, as in the following form

$$\begin{cases} \dfrac{dH}{dt} = rH - aHP \\ \dfrac{dP}{dt} = bHP - mP \end{cases}$$

with the initial conditions

$$t_0 = 0 \ ; \ H(t_0) = H_0 \ ; \ P(t_0) = P_0$$

where *P* is number of predators and *H* represent individs number in prey.

The grows rates are: *r* for *H*, *a* is the rate of predators, *b* is the rate for predators when one individ of prey is reated and *c* is the death rate in predators' individs.

This system is very strong of nonlinearity represent, with constant coefficients becomes surprising with changes them values.

## 2  Nonlinear system Lotka-Voltera with bifurcations and chaos

After many trails with different coefficients will be to obtain bifurcations, so

$r = 0,2$; $a = 0,8$; $b = 1,03$; $m = 0,04$; $\Delta t = 0,25s$; $H_0 = 10$; $P_0 = 2$

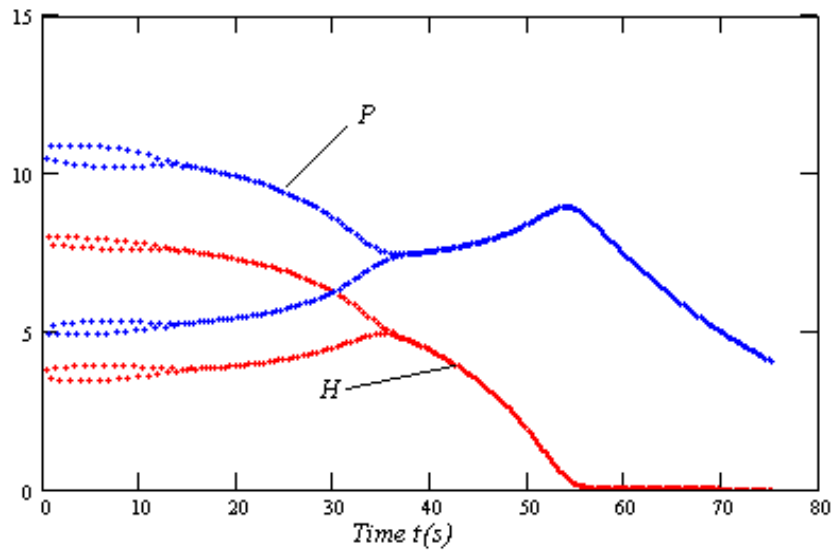

*Figure 2.1* represents diagrams of prey and predators with their bifurcations, at the begining, after this are recombined and the bifurcations become in nonlinear and tendeces to linearity in final of sequences.

The system was integrated by numerical method Runge Kutta or Euler II for any coefficient's values.

## 3 Only simple transformation of nonlinear to linear system

The coefficients leading system to a good representation of linear systems

$r = 1000$; $a = 1000$; $b = 100$; $m = 0.00001$; $H_0 = 1$; $P_0 = 1$

$h := 0.003$

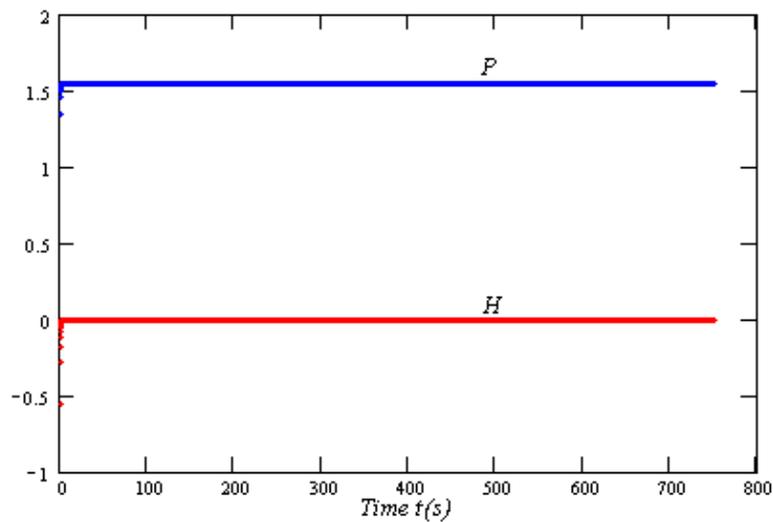

***Figure 3.1*** *represents diagrams of prey and predators with their linearity at the begining to final sequences*

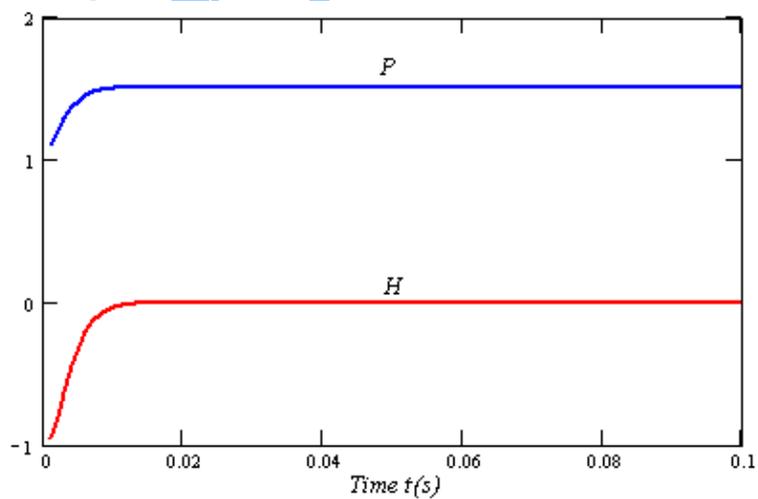

***Figure 3.2*** *represents diagrams of prey and predators with their linearity at the begining to begining sequences*

23

**Conclusions**

The biology system of differential equations utility in the theory is well known and with research in this area is possible many problems to do in action. So, this system becomes linear about any coefficients and especialy with their values. In this way the nonlinear systms have some properties from chaos and destructions to linerity and comands.